# InAs Heteroepitaxy on Nanopillar-Patterned GaAs (111)A

Vinay S. Kunnathully[1,2], Thomas Riedl[1,2], Alexander Trapp[1,2], Timo Langer[1,2], Dirk Reuter[1,2], Jörg K. N. Lindner[1,2]

[1]Paderborn University, Department of Physics, Warburger Straße 100, 33098 Paderborn, Germany

[2]Center for Optoelectronics and Photonics Paderborn (CeOPP), Warbuger Straße 100, 33098 Paderborn, Germany

## Abstract

Heteroepitaxy on nanopatterned substrates is a means of defect reduction at semiconductor heterointerfaces by exploiting substrate compliance and enhanced elastic lattice relaxation resulting from reduced dimensions. We explore this possibility in the InAs/GaAs(111)A system using a combination of nanosphere lithography and reactive ion etching of the GaAs(111)A substrate for nano-patterning of the substrate, yielding pillars with honeycomb and hexagonal arrangements and varied nearest neighbor distances. Substrate patterning is followed by MBE growth of InAs at temperatures of 150 - 350°C and growth rates of 0.011 nm/s and 0.11 nm/s. InAs growth in the form of nano-islands on the pillar tops is achieved by lowering the adatom migration length by choosing a low growth temperature of 150°C at the growth rate 0.011 nm/s. The choice of a higher growth rate of 0.11 nm/s results in higher InAs island nucleation and the formation of hillocks concentrated at the pillar bases due to a further reduction of adatom migration length. A common feature of the growth morphology for all other explored conditions is the formation of merged hillocks or pyramids with well-defined facets due to the presence of a concave surface curvature at the pillar bases acting as adatom sinks.

## 1. Introduction

Growth of lattice mismatched III-V compound semiconductors on nanopatterned substrates is interesting for the fabrication of site-controlled and low-defect semiconductor layers and nanostructures. Among the zinc-blende III-V semiconductors InAs exhibits properties that are attractive for optoelectronic applications, namely a narrow bandgap, a high electron mobility and alloying capability with other III-Vs. Site-controlled growth of InAs and $In_xGa_{1-x}As$ quantum dots and (lateral) nanowires has been achieved on substrates patterned with nanoscale holes or grooves by making use of the high In mobility in conjunction with the surface curvature driven chemical potential landscape [1,2]. When growing heteroepitaxial semiconductors on nanoscale, laterally freestanding substrate areas like the top faces of pillars or ridges, the enhanced elastic strain relaxation aids to reduce the density of misfit-related defects [3]. Moreover, specific arrangements of III-V nanostructures can be realized on such patterned growth areas. For example, Konkar et al. found evidence for the formation of single, double and triple rows of InAs quantum dots on top of GaAs(001) nanomesa stripes which were obtained by size-reducing epitaxy [4]. While some studies analyze InAs heteroepitaxy on nanopatterned GaAs(001), there is a lack of knowledge on InAs growth on nanopatterned GaAs(111)A. The InAs/GaAs(111)A system is characterized by a low binding energy of In adatom sites [5] and slow In-$As_x$ ($x$ = 2, 4) reaction kinetics leading to a high In adatom mobility [6] and preference for 2D layer growth on planar GaAs(111)A [7,8]. These characteristics may provide the ground for the realization of tailored InAs nanostructures on nanopatterned GaAs(111)A.

In the present contribution we explore the influence of MBE growth conditions, i.e. temperature and growth rate, on InAs heteroepitaxy on nanopillar-patterned

GaAs(111)A substrates. The resulting growth morphologies are discussed in terms of the In adatom surface migration length. For the nanopatterning we employ self-organized nanosphere lithography (NSL), which is a low-cost and high-throughput technique, followed by reactive ion etching (RIE).

## 2. Experimental

All experiments were performed on sixths of 3-inch GaAs(111)A wafers procured from MaTeck GmbH. Surface nanopatterning starts with the deposition of mono- and doublelayers of 220 nm diameter polystyrene spheres with 3% coefficient of variation, using an aqueous suspension with 10% solid fraction from Thermo Fischer Inc. Self-organization of beads is initiated by spreading the suspension with a doctor blade on a hydrophilized GaAs substrate. Either mono- or doublelayers of hexagonally stacked sphere layers result from convective self-assembly [9,10] and can be used as shadow mask for deposition of a metallic hard mask. In the present study, triangular or hexagonal sphere mask openings with a pitch of 1.155 *r* or 2.0 *r*, corresponding to mono- or double ayers, respectively, [11] were realised, with *r* being the radius of polymer spheres used. Thermal evaporation of nickel onto the sphere mono- or doublelayers and subsequent dissolution of polymer spheres in tetrahydrofuran resulted in the formation of hexagonal arrays of nickel dots. RIE of the substrate using $SiCl_4$ plasma (5 sccm, 3.5 mTorr, 195 W RF, 55 s in a Plasmalab 100, Oxford Instruments) resulted in the formation of GaAs pillar-patterned substrates with pillar diameters of ≈ 40 and 25 nm for mono- and doublelayer sphere arrays, respectively, and a height of about 90 nm. The mean separation between the nearest-neighbour pillars was $131 \pm 6$ nm and $232 \pm 7$ nm, the larger separation corresponding to doublelayer regions. A detailed description of GaAs nano-pillar array fabrication on GaAs(111)A substrates is given in ref. [12].

In preparation of the heteroepitaxial growth, the nanopatterned substrates were immersed first in diluted $H_2SO_4$ to dissolve the Ni hard mask and then in diluted HF solution to dissolve the surface oxide. Residual surface oxides were removed using atomic hydrogen cleaning for 3 minutes at 350°C in the MBE growth chamber immediately before growth.

The MBE overgrowth of the nanopillar- patterned GaAs with nominally 15 nm InAs (if not stated otherwise) was performed at a constant $As_4$ beam equivalent pressure of approximately $2.5 \times 10^{-5}$ mbar. The objective of MBE growth studies was to understand the effect of growth temperature, growth rate and indium cell inclination on the morphology of InAs formed at pillars with different pillar size and area densities. A summary of growth conditions is presented in Table 1. All InAs growth experiments were performed at a fixed As source flux angle of 40° with respect to the substrate normal, while the In source was placed at inclinations of either 5° or 40° to the substrate normal. The samples were rotated at a speed of 10 rpm to achieve an azimuthal direction-independent exposure of the patterned substrate surface to the $As_4$ and In beams .The chosen parameters of our experiments allow us thus to explore the effect of temperature, growth rate, deposition angle, and GaAs nanopillar arrangement on InAs/GaAs nano-heteroepitaxy.

The growth morphology of the InAs epilayers was characterized using a field-emission scanning electron microscope (Raith Pioneer) operated at 10 kV and transmission electron microscopy (JEOL JEM-ARM200F, $C_s$-probe corrected) at 200 kV in bright-field TEM and high-angle annular dark-field STEM modes. For investigation of the surface morphology of the InAs growth an atomic force microscope (AFM) NanoSurf Mobile S operated in the contact mode was used. For estimation of the area fraction of microstructural features, the open source image analysis software ImageJ was employed.

Table 1. Growth parameters employed in this study. The $As_4$ beam equivalent pressure was maintained at approximately 2.5 × $10^{-5}$ mbar during all experiments. For all samples compiled here the nominal InAs thickness is 15 nm.

| **Sample name** | **Growth parameters** | | | |
|---|---|---|---|---|
| | Substrate temperature (°C) | Growth rate (nm/s) | Indium cell inclination (degrees with respect to substrate normal) | V/III ratio |
| T1 | 150 | 0.011 | 0 | 400 |
| T2 | 350 | 0.011 | 0 | 400 |
| T2' | 350 | 0.011 | 40 | 400 |
| R1 | 300 | 0.011 | 40 | 400 |
| R2 | 300 | 0.11 | 0 | 40 |

**3. Results and discussion**

Nanopillar-patterned areas of approximately 1.3 cm × 2.5 cm were obtained by nanosphere lithography and RIE as described in Section 2. Typical patterns fabricated by using sphere monolayers are shown in Fig. 1. In addition to nanopillars features, arrays contain defects (notably line defects or pattern defects) owing to imperfect sphere arrangement in the NSL step. Such defects can give extra information on the influence of surface morphology on the InAs growth.

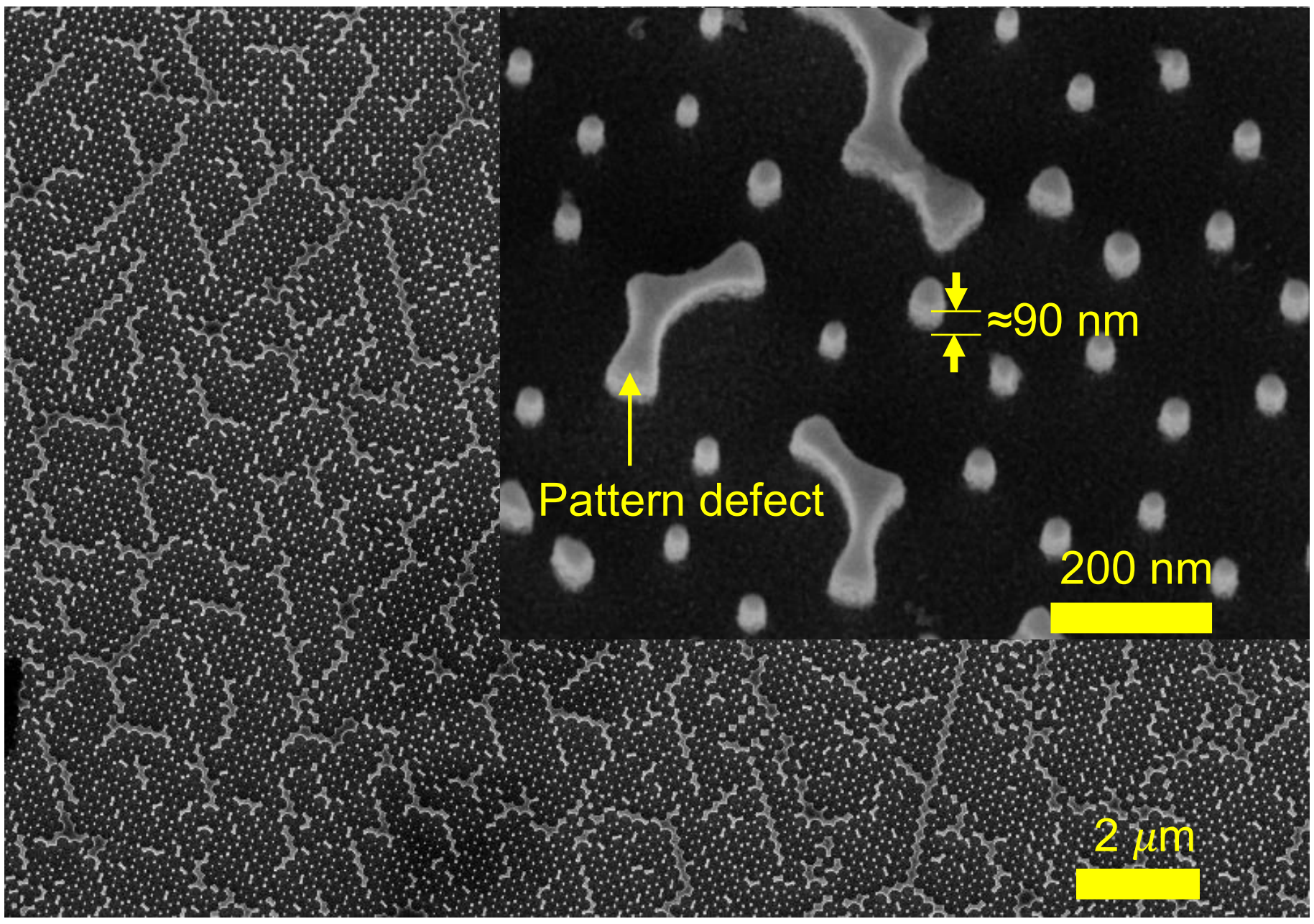

Fig. 1 Top view SEM image of a nanopillar-patterned GaAs surface (obtained from a sphere monolayer). The inset shows a 45° tilted view at higher magnification.

### 3.1 InAs grown on monolayer-patterned areas

Fig. 2 depicts SEM images of InAs overgrown pillar-patterned GaAs obtained from sphere monolayers. A common feature observed is the preferential growth of InAs hillocks around GaAs pillars. In top view, the hillocks show a morphology resembling triangles or triangles with truncated corners, which are also sometimes coalesced (Fig. 2 (a) and (b)). The edges of these triangles are oriented along the $< 1\bar{1}0 >$ directions and the normals to these edges in the (111) plane are oriented along the $< 11\bar{2} >$ directions.Under As rich conditions, the GaAs(111)A surface adopts a 2 $\times$ 2 surface reconstruction, the potential energy surface (PES) of In adatoms on which has minima oriented along the $< 1\bar{1}0 >$ directions [13], which provides a low energy path for Indium adatoms to migrate. This results in a diffusion anisotropy and thus the step edges of InAs hillocks are oriented along the $< 1\bar{1}0 >$ directions.

Formation of such hillocks with edges oriented along the $< 1\bar{1}0 >$ directions is widely reported in literature [14, 15, 16] for homo- and heteroepitaxial growth on fcc (111) surfaces. A high Ehrlich-Schwoebel barrier [17] at step edges of 2D islands restricts the downward motion of adatoms and facilitates 3D growth by increasing their residence time on a 2D terrace and thus also the probability of forming new 2D layers on top of existing ones [18,19]. This effect, together with the diffusion anisotropy and the low energy of {110} type facets is responsible for the formation of the observed hillocks during MBE growth. In the present study, the concave base edges of GaAs nanopillars and line defects act as adatom sinks [20] and are the preferred sites for formation of hillocks, as seen in Fig. 2 (a)-(h). Hence, unlike in most of the studies carried out on planar substrates, in the present study the sites for the formation of hillocks are not random positions. A feature observable on closer examination in the case of samples grown at 300°C and 350°C is the presence of step bunches (Fig. 2 (a), (b) and (c)), which indicates that growth is mediated by the step flow growth mechanism.

The effect of InAs growth rate on growth morphology is considered first on the basis of images in Fig. 2 (b), (f) and (c), (g). The most noticeable difference between Fig. 2 (b) and (c) is the reduced degree of coalescence of adjacent hillocks at the higher growth rate (Fig. 2 c). In this case, the higher In adatom flux increases the nucleation rate and as a consequence reduces the migration length on account of the increased probability of adatom capture by newly formed islands [17] on the hillock and also on the planar substrate between pillars. In the present case, further InAs growth takes place preferably on the widely relaxed surfaces of the existing InAs hillocks and at the kink and steps present on the pillar sidewalls. We also observe that all the hillocks are oriented in the same direction throughout the sample for all samples reported here.

Now let us consider the temperature dependence and the influence of deposition angle. Merged hillocks with step bunches on their surface (Fig. 2(a)) and well-defined facet surfaces (Fig. 2 (e)) are observable in the case of a sample grown at 350°C (Fig. 2 (e), sample T2). The InAs growth morphology of a sample grown at the same substrate temperature but with an inclined In cell (sample T2’) is the same as for normal incidence of In. A SEM top view image for inclined deposition is shown in the Supplementary Materials section S1, Fig. S1 (a). There is no pronounced morphological difference between samples grown at 350 and 300°C (Fig. 2 (a), (e) and (b), (f), respectively). In comparison, at 150°C (Fig. 2 (d), (h)) the edges of hillock’s facets look irregular and no signatures of step bunching are observable. Yet another difference is visible, i.e. the presence of numerous small islands on the planar substrate areas and the formation of nano-islands on the pillar tops of this sample.

These observations reflect the role of growth temperature in regulating the migration length which decreases exponentially with temperature. In addition to this the incorporation rate of As is strongly enhanced by a reduction of GaAs(111)A substrate temperature [21]. This also enhances the nucleation rate and promotes the formation of InAs islands on the planar substrate surfaces. For all other temperatures above 150°C examined in this study, the pillar tops were observed to be devoid of InAs islands (also confirmed using cross-sectional TEM, not shown here).

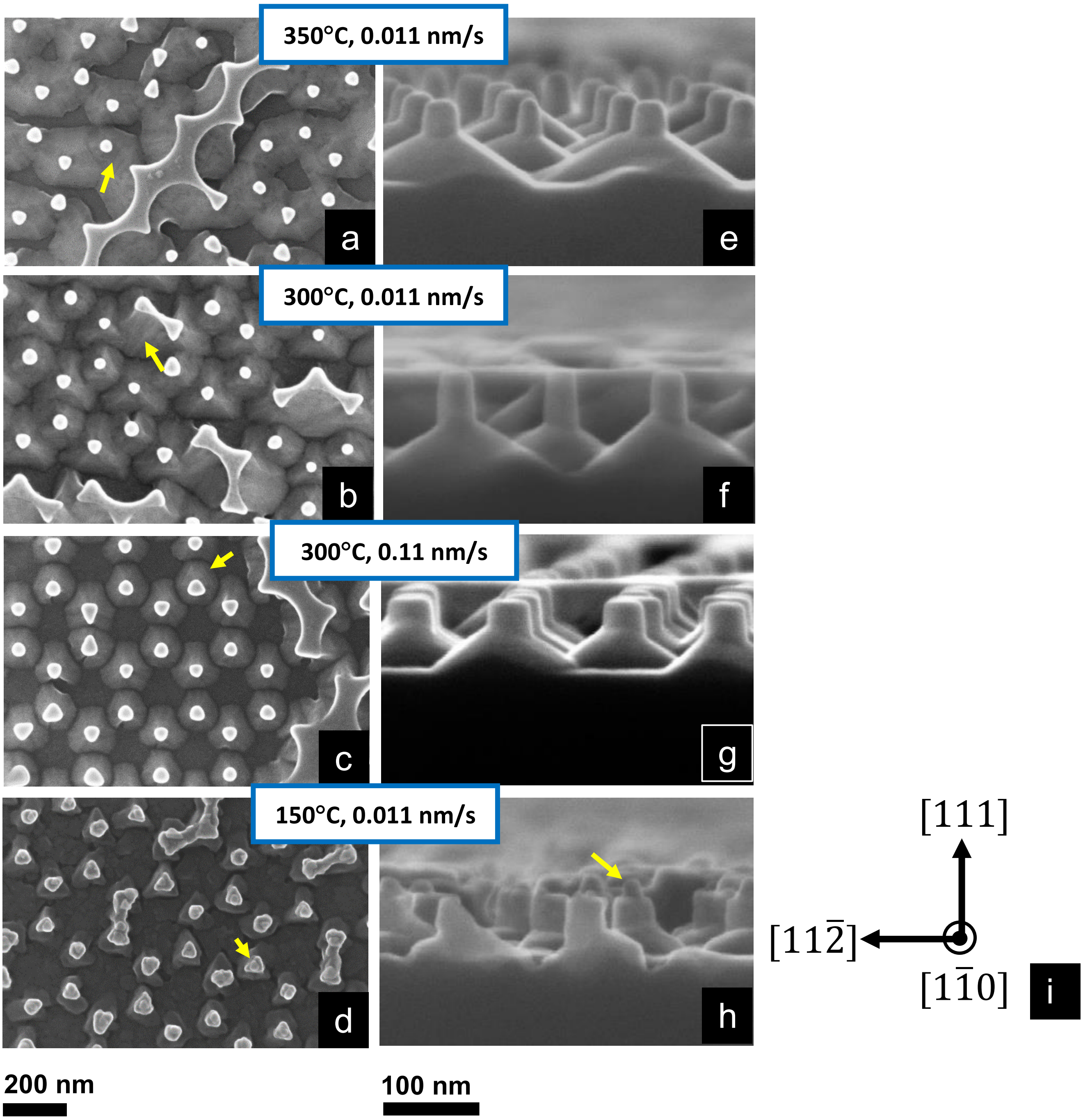

Fig. 2 SEM images of nanopillar-patterned GaAs surfaces overgrown with InAs at different conditions in plane view (a-d) and cross sectional view (e-h). All surface patterns shown here were created using NSL monolayers. The yellow arrows in (a) and (b) point at step bunches. In (d) and (h) the yellow arrows point at InAs nano-islands. Crystallographic directions in the cross-section images (e) – (h) are given in (i).

An estimate of the dimensions of InAs islands was made using bright-field TEM images such as in Fig. 3 (a). The average diameter and height were found to be 21 nm and 15 nm, respectively. The bright-field image in Fig. 3 (a) and the HAADF-STEM image in Fig. 3 (b) also reveal the growth of InAs on the pillar side walls as well as at the base of the pillars. An analysis of HAADF-STEM images from GaAs nano-pillars overgrown with InAs at the same conditions as sample T1 but with lower deposited InAs thickness (2 nm) reveals the presence of a thin InAs wetting layer (Information presented in Supplementary Materials section S2). The detection of such a wetting layer points to the possibility of InAs growing on top of GaAs pillars via a Stranski-Krastanov type of mechanism.

The projected area fraction covered with InAs for the samples compiled in Table 1 is represented in Fig. 4. The projected areas were estimated by manually coloring and thresholding top view SEM images like in Fig. 2(a-d) to enable the estimation of area fractions occupied with InAs in hillocks. This manual approach was chosen to allow for an exclusion of defective pattern areas in the analysis. Defect free monolayer patterned regions with a minimum size of 0.3 $\mu m^2$ were used to estimate the average projected area fraction of InAs. The sequences involved in such image processing are outlined in the Supplementary Materials section S3.

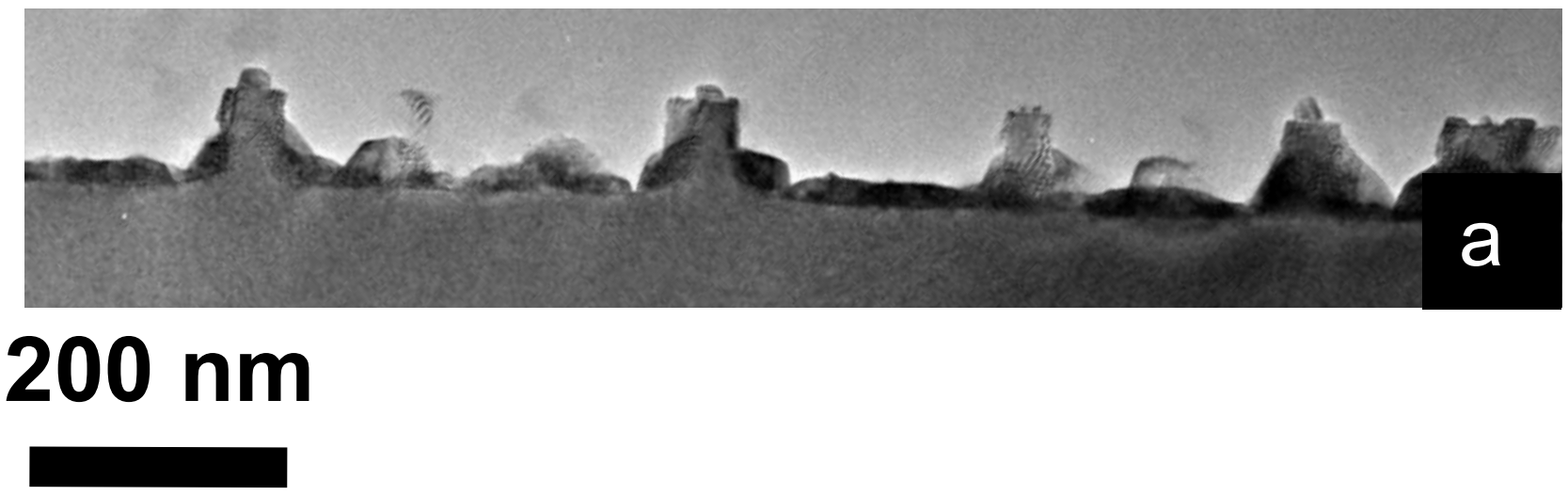

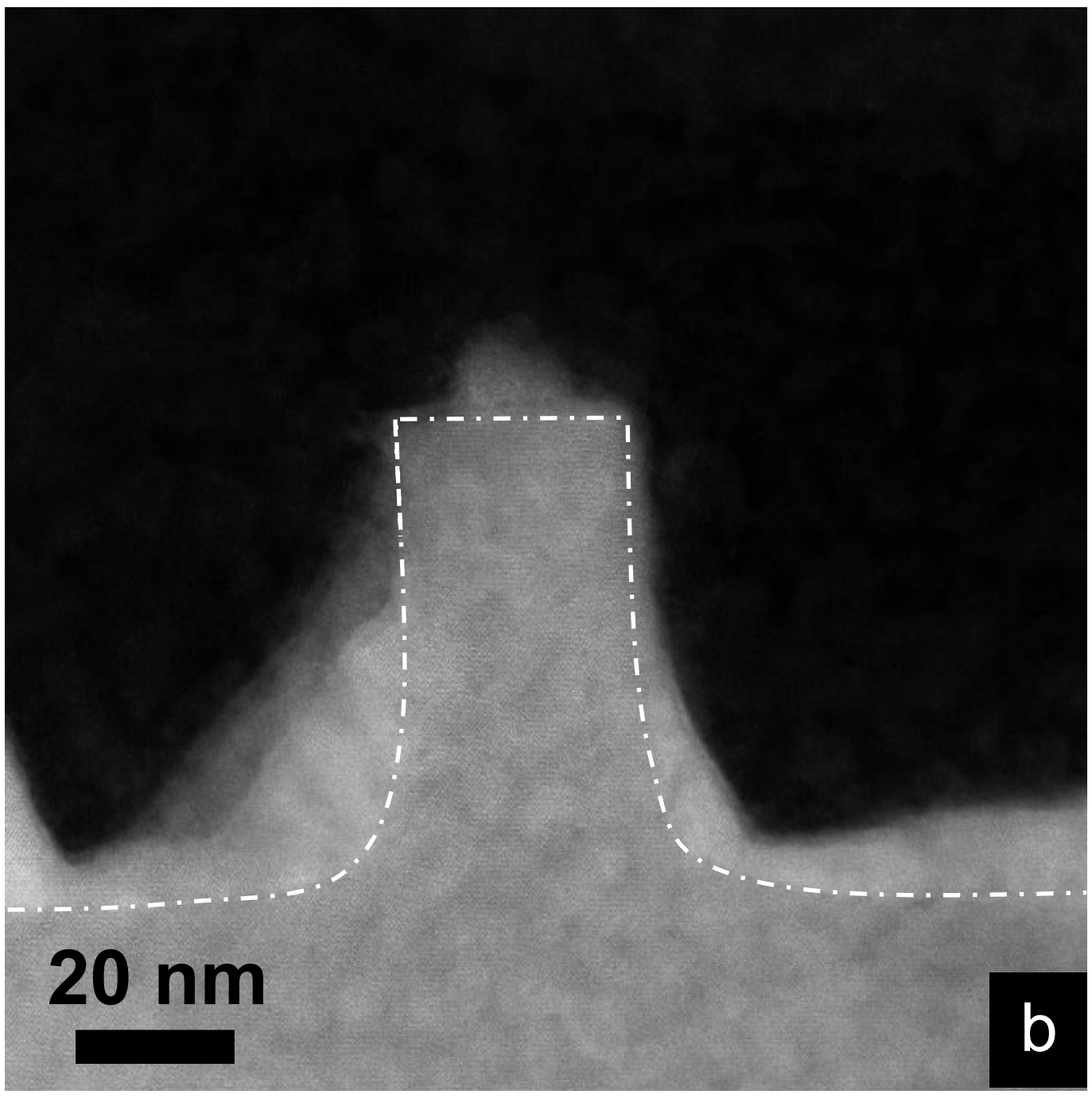

Fig. 3. (a) Cross-sectional TEM bright-field image of a nanopillar-patterned GaAs surface overgrown with InAs at 150°C (sample T1) (b) HAADF-STEM image of a single pillar of the same sample at higher magnification. The white dotted line indicates the interface between InAs and GaAs.

At least five such areas excluding line defects and containing a minimum of 360 pillars were analyzed to compute the standard deviation shown as vertical bars in the plot of Fig. 4. Fig. 4 reveals a decline of the projected area fraction of InAs hillock coverage with an increase in growth temperature for samples grown with a constant growth rate of 0.011 nm/s. Such a trend is the result of enhanced adatom migration towards the bases of pillars and line defects in the pattern, which reduces the amount of InAs growing in the free areas between the pillars. The data points (shown in red) corresponding to samples T2 and T2' with identical temperature of 350°C but different In effusion cell inclination angles show an almost perfect coincidence. This suggests that at least at the highest temperature the growth morphologies depend on growth kinetics rather than on the direction of adatom arrival. A difference of 26% in InAs coverage is observed between the two samples grown at 300°C at different growth rates. This is a result of differences in lateral migration lengths. A 100% coverage is found for the growth at 150°C, corroborated by morphological observations in Fig 2. (d) and (h).

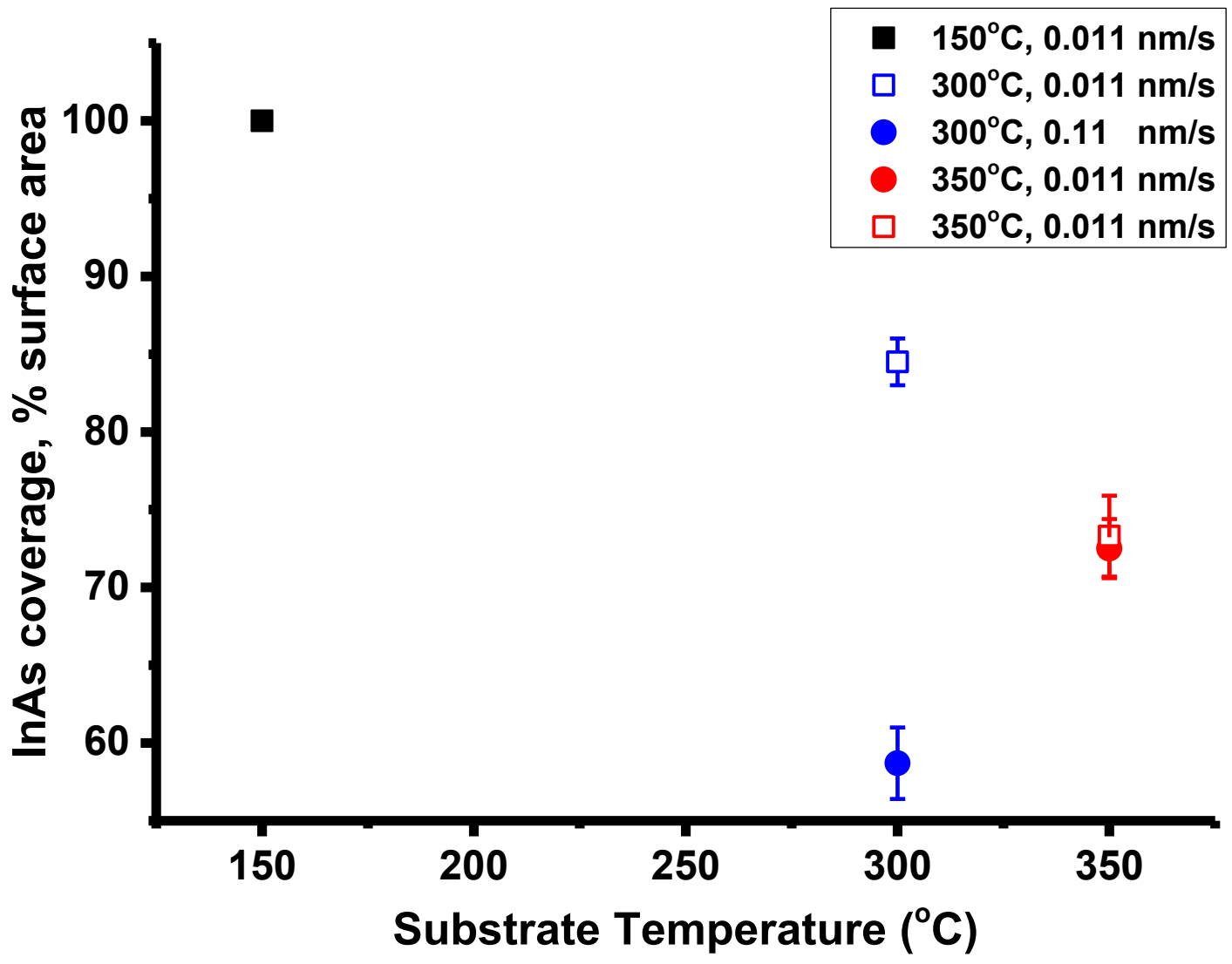

Fig. 4 The percentage of projected surface area covered with InAs hillocks in patterned areas obtained using sphere monolayers as a function of growth temperature. The square open points indicate that an inclined In effusion cell was used whereas round closed points indicate usage of the vertical cell.

### 3.2 InAs grown on double layer patterned areas

Pillar arrays fabricated using doublelayer NSL masks instead of monolayer masks are characterized by an area density of pillars reduced by a factor of two with respect to pillar arrays based on monolayers and by reduced pillar diameters of 25 instead of 40 nm. As for the case of higher pillar densities, again InAs pyramids with edges oriented along the $<1\bar{1}0>$ directions at their bases are observed to grow around GaAs pillars (Fig. 5). These InAs structures are distinct from hillocks formed on monolayer patterned areas in the sense that they show well-defined edges, three major and six minor facets as shown schematically in Fig. 6 (d). Facets are particularly well visible on samples grown at temperatures of 350°C and 300°C (Fig. 5 a,b). In case of the 300°C high-growth rate sample R2, the pyramid edges appear to be rounded (Fig. 5c). Sample R2 also shows on closer observation the presence of flat InAs islands surrounding the pyramids. Such islands are a result of the high nucleation rate and this should slightly reduce the volume of the InAs pyramids due to fewer adatoms available. In conjunction with higher pyramid sidewall inclination (shown below), this also prevents adjacent InAs pyramids from coalescing with one another as observed in case of R1 and T2 (Fig. 5 (a), (b)). At 150°C (sample T1) InAs growth occurs in a very irregular way (Fig. 5 (d)). Pillars are surrounded by numerous irregularly shaped InAs islands. On top of the pillars the formation of InAs nano-islands is observed (Fig. 5 (d)), similar as for the growth on larger diameter pillars with higher density in section 3.1 (Fig. 2 (d), (h)). Such a result is promising for the formation of InAs nano-islands on smaller diameter pillars at well-defined substrate positions

AFM surface profiles of the pillars were analyzed quantitatively to index the facets formed on the pyramid surfaces (Fig. 6 (a), (b)). A typical AFM height profile of a major facet corresponding to a pyramid formed at 350°C (sample T2) is shown in Fig. 6 (c). AFM height profiles of the major facets along the $<11\bar{2}>$ directions yield facet inclination angles of 35.5±1.4°, 36.2±1.7° and 38.9±1.7° for samples T2, R1 and R2 respectively. The measured angles of samples T2 and R1 are close to the theoretical value of 35.26° between the $\langle 110 \rangle$ facet normals and the [111] plane normal of the substrate. The same low energy {110} facet orientation has also been reported in the literature for InAs and GaAs grown on GaAs(111)A surfaces [22, 23, 24]. The higher growth rate sample R2 displays a larger facet inclination angle than samples T2 and R1, which can be ascribed to the enhanced second-layer nucleation rate and 3D like growth for higher group III flux and lower V/III ratio [17].

For the minor facets, inclination angle measurements with respect to the substrate surface yield values of 31.2°±1.8° and 39.6°±1.8° with respect to the substrate normal. However, since the intercepts line of the minor facets with the substrate show a wide range of variability (Fig. 5 (b), Fig. 6 (b)) the facet normals cannot be unequivocally assigned.

Pyramid morphologies of samples T2 and T2' with identical growth temperature but different In deposition angles are identical as shown in Supplementary Data file (Fig. S1 (a), section S1.). Models based on bulk atomic configurations at step edges are used to explain the pyramidal morphology of deposited materials during epitaxial growth. Despite neglecting any surface reconstructions, these models are able to explain the formation of pyramids on fcc (111) surfaces [24-26]. There are six $<11\bar{2}>$ type directions lying in the GaAs(111)A plane, which are normal to the step edges parallel to $<1\bar{1}0>$ directions. Among these six, three ($[2\bar{1}\bar{1}]$, $[\bar{1}\bar{1}2]$ and $[\bar{1}2\bar{1}]$) are of A type and three ($[11\bar{2}]$, $[1\bar{2}1]$ and $[\bar{2}11]$) are B type step edges. Type A step edges are characterized by two dangling bonds per edge atom whereas it is only one in case of type B type step edges. This asymmetry results in a preference of arriving adatoms to be incorporated in type A step edges. This ultimately leads to faster growth and shrinking of type A step edges, resulting in the formation of a triangular base of the pyramids. As this basis is identical for each pyramid, the pyramids all show the same orientation.

Facet analysis of the InAs pyramids shows that the three major facets are of {110} type, each two of them are separated by two minor facets, which are likely to correspond to higher index planes due to their smaller projected area [27] compared to the major facets. The formation of such pyramidal structures with major and minor facets is a result of the system adopting an energetically favorable configuration under non-equilibrium processing conditions such as the MBE growth [28]. Another general observation of InAs growth on double layer patterned areas is the reduced lateral coalescence of the hillocks due to the larger separation of the pillar bases. This also shows that the pyramids are the preferred shape in the absence of coalescence with adjacent growth fronts of neighboring InAs hillocks.

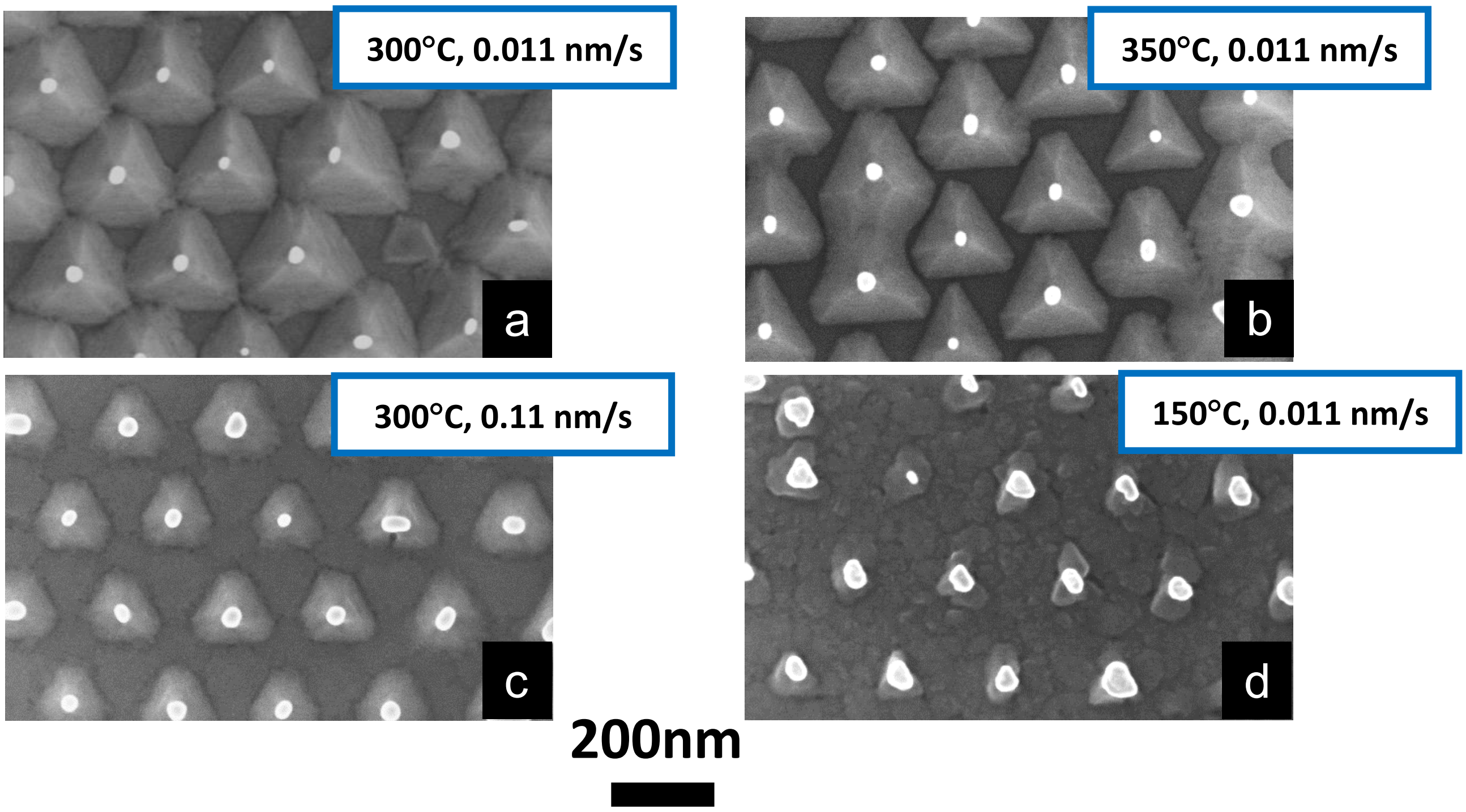

Fig. 5 SEM top view images of InAs pyramids grown around GaAs nanopillars fabricated using sphere doublelayer masks for (a) sample R1, (b) sample T2, (c) sample R2, and (d) sample T1.

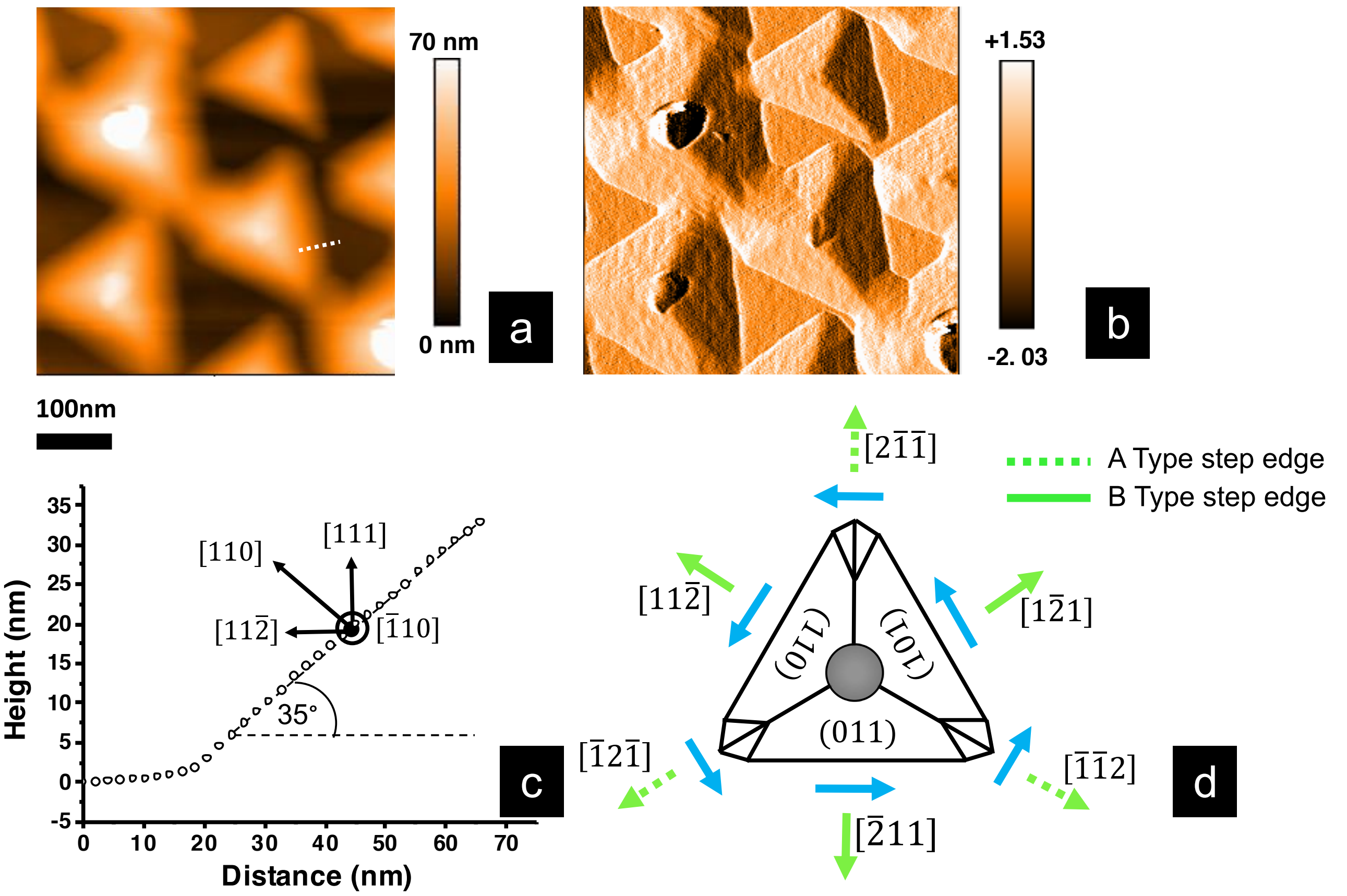

Fig. 6 AFM maps of InAs pyramids on sample T2 grown at 350°C and 0.011 nm/s growth rate. (a) Raw height data, (b) derivative plot of (a) and (c) height profile along the dotted line in (a). (d) Schematic of relevant crystallographic directions and morphology of InAs pyramids.)

## 4. Conclusions

The growth morphology resulting from heteroepitaxial growth on nanopatterned substrates can be tailored by adjusting adatom mobility and spatial distribution of adatom sinks. In the present study, we have identified growth conditions which enable the formation of InAs quantum dots on GaAs (111)A pillar tops by a Stranski-Krastanov type growth mechanism and ring-like nanostructures at pillar bases by tuning adatom mobility. Formation of well-defined pyramidal InAs structures at pillar bases shows the additional degree of freedom provided by variation of nearest neighbor distance of adatom sinks under conditions of high adatom mobility. These insights could pave way for site-selective growth of interesting InAs/GaAs nanostructures.

**Acknowledgements**

The authors acknowledge the financial support provided by German Science Foundation (DFG) under the project numbers Ri2655/1-1 and Li449/16-1. We also thank Carl Zeiss AG, Oberkochen for preparation of a TEM FIB lamella.

**Supplementary Material**

# InAs Heteroepitaxy on Nanopillar-Patterned GaAs (111)A

Vinay S. Kunnathully[1,2], Thomas Riedl[1,2], Alexander Trapp[1,2], Timo Langer[1,2], Dirk Reuter[1,2], Jörg K. N. Lindner[1,2]

[1]Paderborn University, Department of Physics, Warburger Straße 100, 33098 Paderborn, Germany

[2]Center for Optoelectronics and Photonics Paderborn (CeOPP), Warbuger Straße 100, 33098 Paderborn, Germany

## S1 SEM images of mono and double layer patterned areas of sample T2´

SEM top view images of sample T2’ corresponding to (a) mono- and (b) doublelayer regions. Fig. S1 (a) and (b) are morphologically identical to those of sample T2 (Fig. 2 (a) and Fig. 5 (b) shown in the main body of this article)

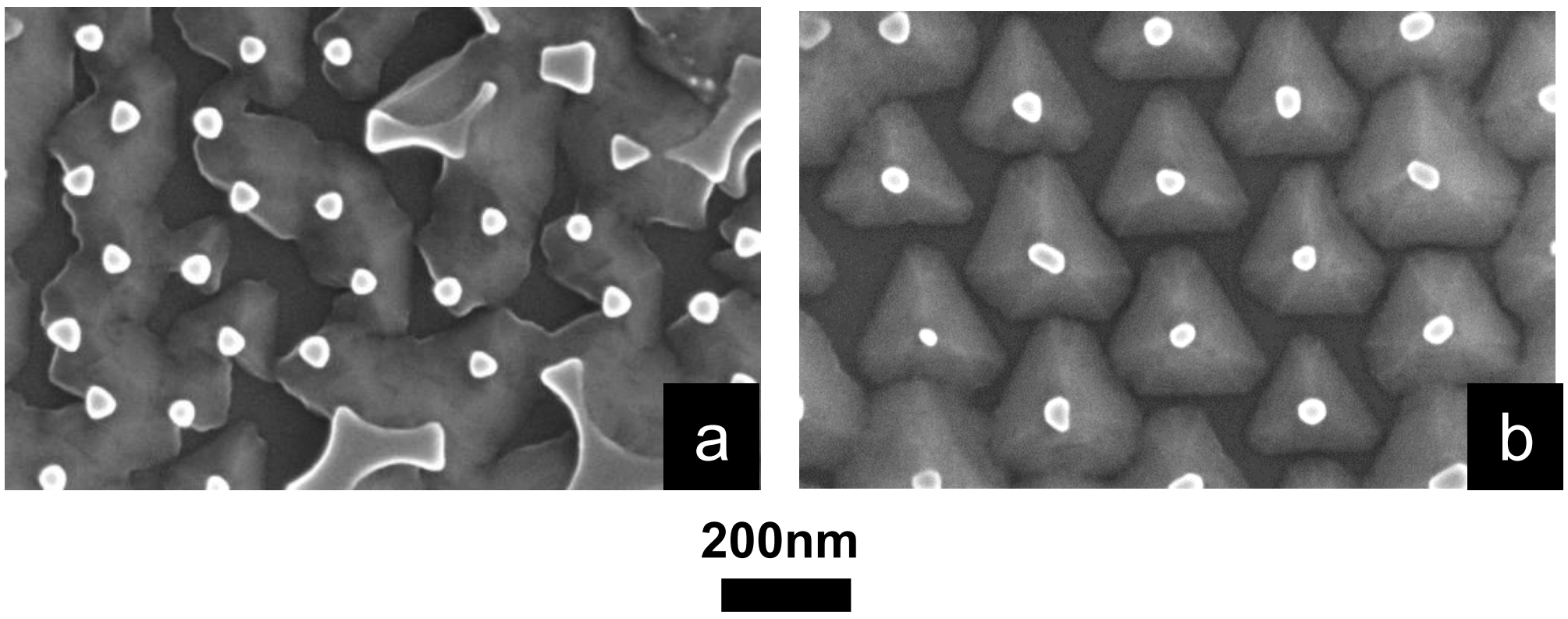

Fig. S1 Top view SEM image of sample T2’ corresponding to a) monolayer and b) doublelayer patterned areas, respectively.

## S2. Detection of InAs wetting layer

A high-resolution STEM-HAADF image of a sample grown at similar conditions as sample T1 (150°C, 0.011 nm/s) but with the thickness of deposited InAs of only 2 nm is shown in Fig. S2-1. The sample is in $[\bar{1}10]$ zone axis orientation. Image (b) displays an InAs island grown on top of the GaAs nano-pillar shown in Fig. S2-1 (a).

The intensity profile measured in the region of interest marked in Fig. S2-1 (b) in orange color is plotted in Fig. S2-1 (c). The intensity plot in Fig. S2-1 (c) shows enhanced intensity in the locations between the red arrows, also marked by red arrows in Fig. S2-1 (b). The enhanced intensity indicates that an InAs wetting layer is present since the intensity of HAADF-STEM intensities increase proportional to the projected atomic number $Z^{\sim 1.8}$ of the scattering atoms. Note the presence of amorphous material on top of the GaAs pillar, most likely due to contamination or oxidation.

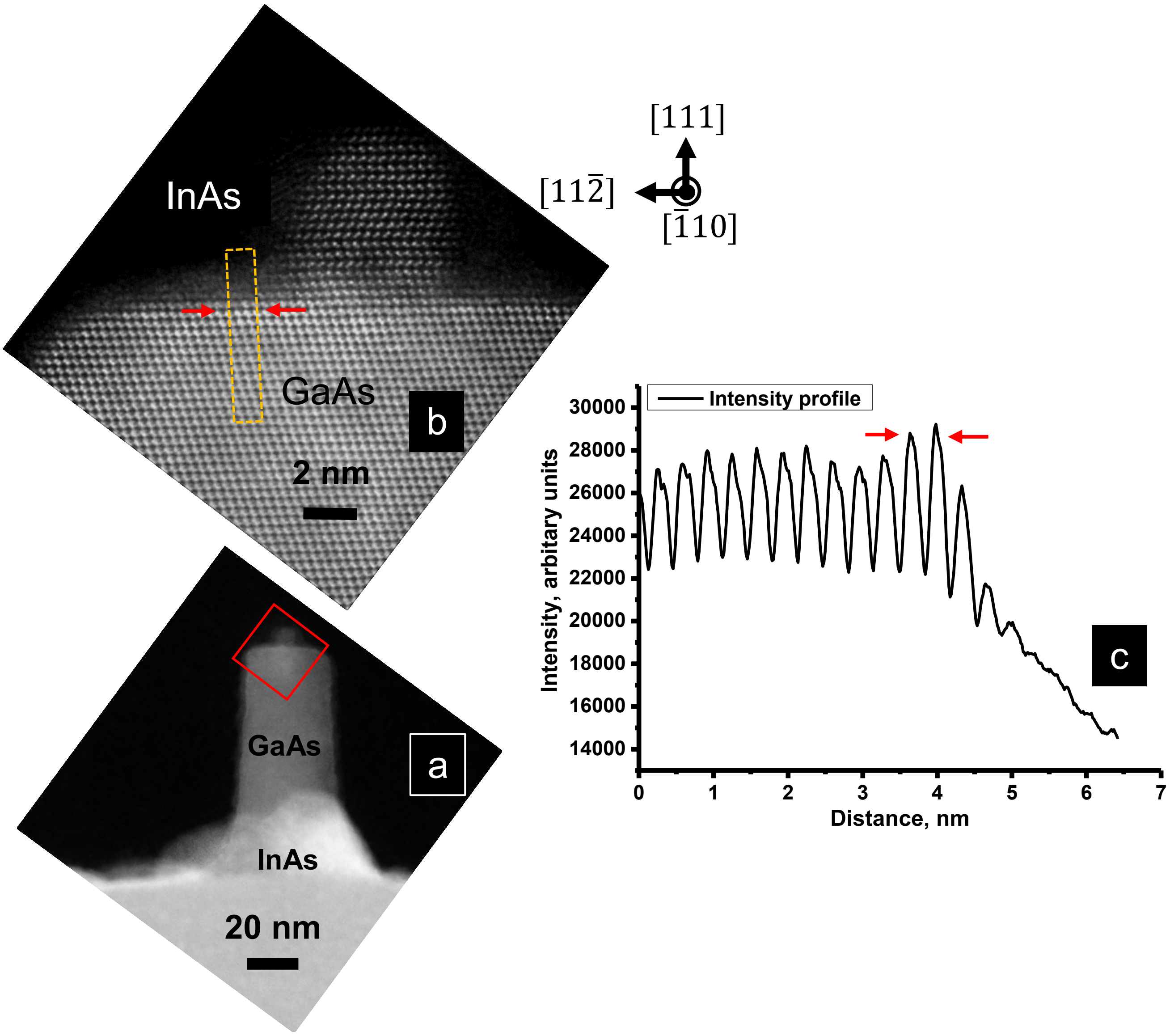

Fig. S2-1 The intensity profile corresponding to the region of interest marked in (b) (orange rectangle) is plotted in (c). Image (b) is a magnified view of an InAs island growing at a GaAs nano-pillar shown in (a) The directions indicated in the figure correspond to specimen orientation for (a) and (b).

HAADF-STEM measurements performed on a GaAs nano pillar after InAs growth corresponding to sample T2’ is shown in Fig. S2-2(a), together with an EDS (energy dispersive X-ray spectroscopy) line scan (Fig. S2-2(b)) along the dotted yellow line in Fig. S2-2(a). The elemental distribution of Indium, Gallium and Arsenic along the line are depicted. The spike in the Indium concentration at the edges of the pillar as seen in Fig. S2-2 (b) is an indication of an Indium rich layer on top of GaAs pillar.

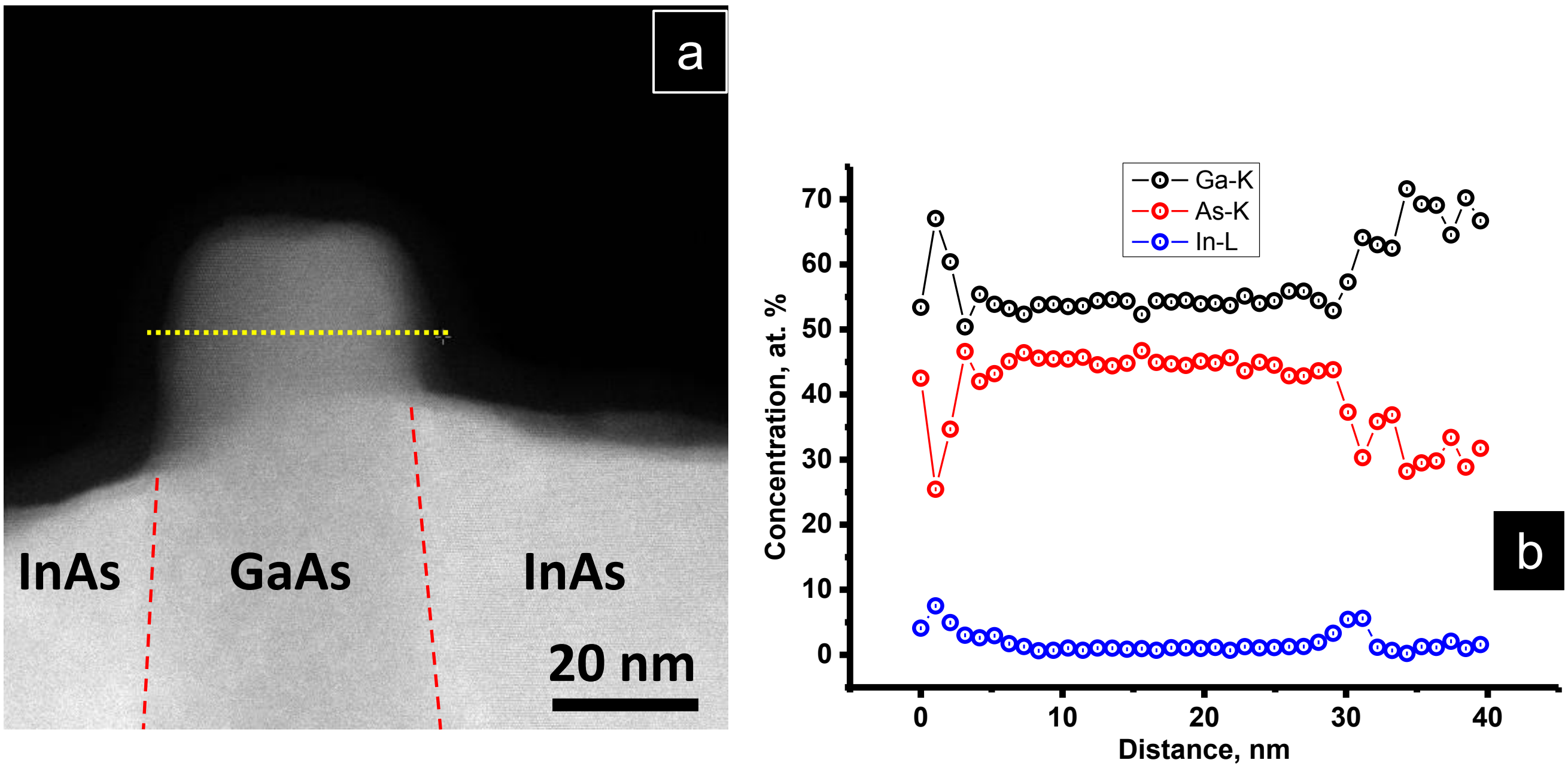

Fig. S2-2 (a) STEM-HAADF image of a GaAs pillar overgrown with InAs.(b) Plot of the elemental concentrations of Ga, As and In determined by EDS along the yellow line in (a).

The evidences presented in Fig. S2-1 and S2-2 are proof for the presence of an InAs wetting layer for all growth temperatures considered in this work.

## S3. Steps involved in determination of InAs area coverage

The sequence of operations performed in the estimation of InAs coverage on nano-pillar patterned substrates is shown in Fig. S3. These operations yield a binary image, from which the area fraction is estimated by image analysis software, ImageJ.

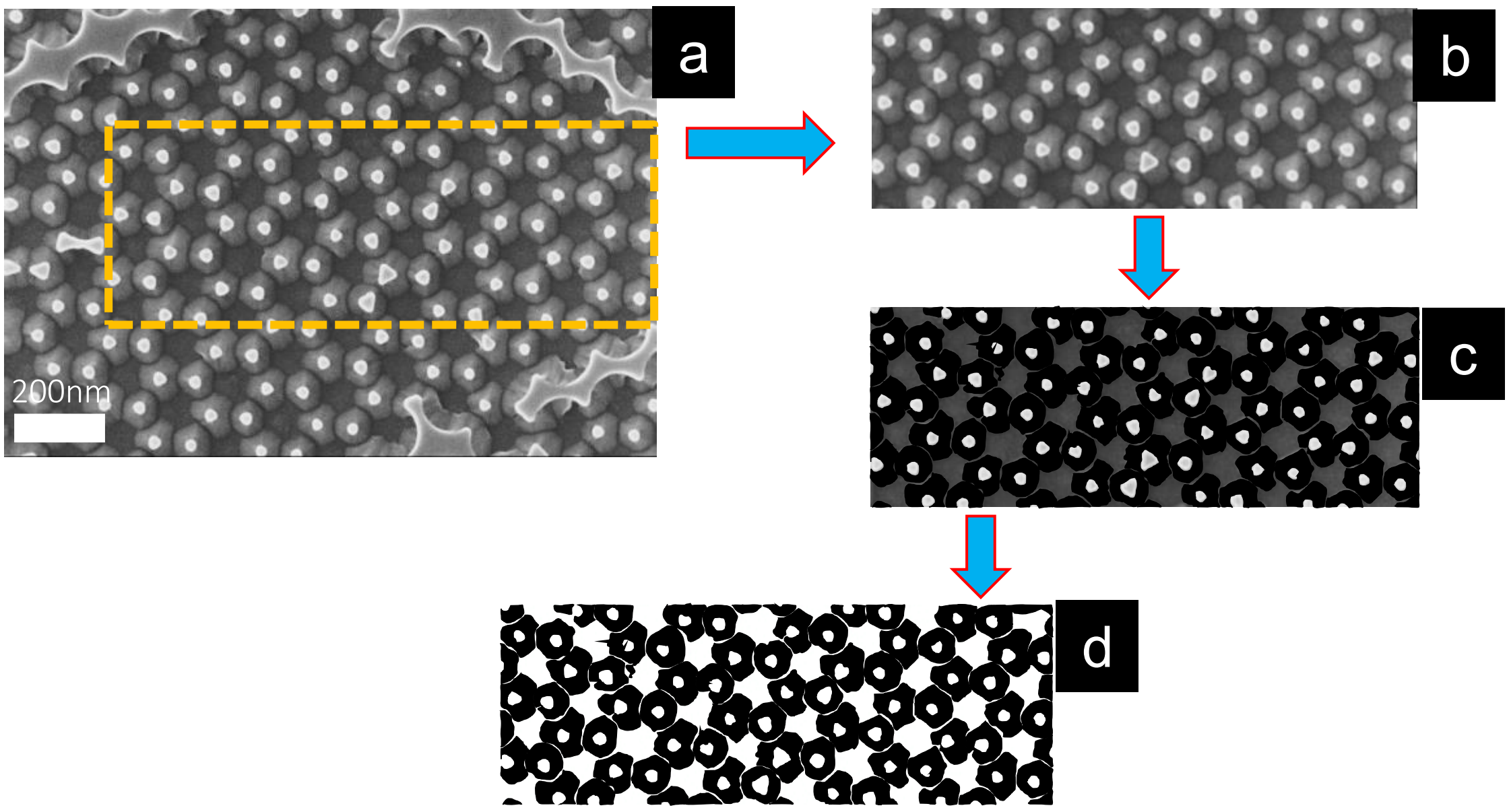

Fig. S3 Top view SEM image of a monolayer region of sample R2 (a) the orange rectangle shows the defect free ROI. (b) Cropped ROI. (c) ROI after manually colouring InAs regions. (d) Binary image obtained after thresholding of (c).